# Multiband superconductivity with sign-preserving order parameter in kagome superconductor $CsV_3Sb_5$

Han-Shu Xu[1], Ya-Jun Yan[1*], Ruotong Yin[1], Wei Xia[4,5], Shijie Fang[1], Ziyuan Chen[1], Yuanji Li[1], Wenqi Yang[1], Yanfeng Guo[4], Dong-Lai Feng[1,2,3*]

[1]*Hefei National Laboratory for Physical Sciences at the Microscale and Department of Physics, University of Science and Technology of China, Hefei, 230026, China*

[2]*Collaborative Innovation Center of Advanced Microstructures, Nanjing, 210093, China*

[3]*Shanghai Research Center for Quantum Sciences, Shanghai 201315, China*

[4]*School of Physical Science and Technology, ShanghaiTech University, Shanghai 201210, China*

[5]*ShanghaiTech Laboratory for Topological Physics, ShanghaiTech University, 201210 Shanghai, China*

**Corresponding author**：<u>yanyj87@ustc.edu.cn; dlfeng@ustc.edu.cn.</u>

**Abstract**

**The superconductivity of a kagome superconductor $CsV_3Sb_5$ is studied by scanning tunneling microscopy/spectroscopy at an ultralow temperature with high resolution. Two kinds of superconducting gaps with multiple sets of coherent peaks and residual zero-energy density of states are observed on both half-Cs and Sb surfaces, implying multiband superconductivity with gap nodes. Sixfold star-shaped magnetic vortex is observed with conventional Caroli–de Gennes–Matricon bound states inside. Magnetic impurities suppress the superconductivity, while non-magnetic impurities do not, suggesting the absence of sign-change in the superconducting order parameter. Moreover, the interplay between charge density waves and superconductivity differs on various bands, resulting in different density of state distributions. Our study provides critical clues for further understanding the superconductivity and its relation to charge density waves in $CsV_3Sb_5$.**

Owing to the special geometry, materials with kagome lattice may possess geometric frustration, flat band and nontrivial band topology, making it an intriguing playground to explore exotic quantum phenomena, such as quantum spin liquid, charge density wave (CDW), Dirac/Weyl semimetal and unconventional superconductivity [1-5]. Recently, a new family of quasi two-dimensional kagome metals $AV_3Sb_5$ (A=K, Rb, Cs) has been discovered, consisting of perfect kagome lattice of V atoms coordinated by Sb [6]. Non-magnetic $AV_3Sb_5$ goes through a CDW transition below $T_{CDW}$ ~ 78-103 K, and then enters the superconducting state below $T_C \sim 0.9$-3.5 K [6-9]. Multiband structure with several Dirac-cone-like bands, flat bands, and topological nontrivial surface states close to the Fermi level, was predicted by previous theoretical calculations and then confirmed by angle-resolved photoemission spectroscopy (ARPES) studies [6-10]. Scanning tunneling microscopy (STM) study found that the CDW order is topological nontrivial with chiral anisotropy [11], which may be responsible for the observation of large anomalous Hall effect [12,13].

As for the superconducting state, unconventional superconductivity is indicated by the possible experimental signatures of spin-triplet paring and an edge supercurrent in Nb/$K_{1-X}V_3Sb_5$ devices and double superconducting domes under pressure [14-18]. Ultra-low temperature thermal conductivity measurements suggest nodal superconductivity in $CsV_3Sb_5$ [15], although magnetic penetration depth experiments point to a nodeless superconductivity [19]. More intriguingly, topological nontrivial surface states were predicted, and thus topological superconductivity was suggested due to the superconducting proximity effect [7,8,10]. The possible appearance of Majorana zero mode was argued as well [20]. These studies imply rich physics in the superconducting state of $AV_3Sb_5$, but our knowledge is still limited, and some of the results are still controversial. In this paper, we systematically studied the superconducting state of $CsV_3Sb_5$ with an ultra-low temperature STM of high energy and spatial resolution, and provide direct evidence and critical clues on the superconducting pairing symmetry, such as the existence of gap nodes, the absence of sign change in the superconducting order parameter, multi-band superconductivity and its distinct interplay with CDW orders.

$CsV_3Sb_5$ single crystals were grown by the self-flux method [6]. Temperature dependence of magnetic susceptibility and resistance were measured using Quantum Design Magnetic Property Measurement System (MPMS) and Physical Property Measurement System (PPMS), and show a superconducting transition temperature of about 2.6 K (Fig. S1 in supplementary materials (SM, Ref. [21])). The STM experiments were conducted in a dilution-refrigerator-equipped STM system with a base temperature of 30 mK and a maximal magnetic field of 15 T along the vertical direction. The effective electron temperature ($T_{eff}$) at the lowest lattice temperature (~30 mK) was calibrated to be 170 mK, as discussed in Fig. S2 in SM. $CsV_3Sb_5$ single crystals were cleaved in high vacuum at about 80 K and immediately transferred into the STM module. PtIr tips were used after being treated on a clean Au (111) surface. The $dI/dV$ spectra were collected using a standard lock-in technique with modulation frequency f = 477 Hz. All experimental data is collected at about 40 mK.

CsV$_3$Sb$_5$ crystallizes in a layered structure of V-Sb sheets intercalated by a hexagonal Cs layer, as shown in Fig. 1(a). The V-Sb sheets consist of a perfect V kagome sublattice, a hexagonal Sb sublattice centered on each kagome hexagon and a honeycomb Sb sublattice below and above each kagome layer. After cleavage, two kinds of terminated surfaces, a half-Cs surface and a honeycomb Sb surface, were observed in our STM study. Fig. 1(b) shows the typical STM image of a half-Cs surface, Cs atoms form a 1 × 2 reconstruction accompanied by a charge modulation with 2$a_0$ period along the stripes. Fig. 1(c) shows the surface morphology of the Sb surface, where the randomly distributed bright protrusions are likely residual Cs atoms, and the dark spot indicated by the white arrow is a Sb vacancy. Atomically resolved STM image in Fig. 1(d) exhibits clear charge modulations, one is a 2$a_0$×2$a_0$ superstructure propagating along all three lattice directions, the other one is a unidirectional 4$a_0$ modulation. The lattice reconstruction and charge modulation on both surfaces can be resolved more clearly in the fast Fourier transform (FFT) images shown in the insets of Figs. 1(b) and 1(c), peaks of $q_{2a}$ and $q_{4a}$ are clearly presented in addition to the atomic Bragg peaks $q_a$. These results are similar to previous STM studies on AV$_3$Sb$_5$ (A=K, Rb, Cs), and were assigned to multiple CDW orders [11,20,22-24].

As shown in Figs. 1(e) and 1(f), two kinds of superconducting gaps, a V-shaped one and a U-shaped one, are observed on both the half-Cs and Sb surfaces, coexisting with the above mentioned CDWs. The V-shaped superconducting gap on the Sb surface in Fig. 1(f) exhibits two pairs of distinct coherence peaks located at $\Delta_1$ = 0.48 meV and $\Delta_2$ = 0.36 meV, and its gap depth (defined as 1 minus the ratio of density of states (DOS) at zero energy to that at 1.5 meV) is about 75%~80%; while the U-shaped superconducting gap has a pair of broad coherence peaks at about $\Delta_3$ = 0.38 meV and a rather flat gap bottom with the gap depth of about 90%. For the half-Cs surface, similar V-shaped and U-shaped superconducting gaps are observed as shown in Fig. 1(e), but the gap sizes differ slightly ($\Delta_1$ = 0.57 meV, $\Delta_2$ = 0.30 meV and $\Delta_3$ = 0.45 meV here). The existence of residual DOS is intrinsic of CsV$_3$Sb$_5$ and is not caused by the instrument-related offsets, as discussed in Fig. S3 in SM. Regardless of Sb surface or half-Cs surface, the two kinds of superconducting gaps are observed in regions with similar STM topography and broad-energy-range spectroscopic features (Fig. S4 in SM). Considering that CsV$_3$Sb$_5$ has multiple Fermi pockets with different orbital characteristics [6-10], these superconducting gaps may originate from different energy bands, which are observed due to the likely different tunneling matrix element effects of the STM tips. Compared with the Sb surface, the energy bands of the half-Cs surface shift downward about 10 ~ 20 meV (Fig. S4 in

SM), suggesting higher electron doping, which may be responsible for the size change of the superconducting gaps on the half-Cs surface.

The distinct lineshapes of the two superconducting gaps, especially near the gap bottom, imply different superconducting gap functions in different energy bands. More intriguingly, a sizable residual DOS exists at the Fermi level for both kinds of superconducting gaps under such a low effective electron temperature of 170 mK (Fig. S2 in SM); and this is extraordinary for the U-shaped gap with a flat gap bottom that usually characterizes a full gap had the gap depth touched 100%. The residual DOS at Fermi level directly verifies finite quasiparticle excitations, which confirms the existence of gap nodes, as suggested in previous thermal conductivity measurements [15]. Furthermore, considering the particular lineshapes of the two types of superconducting gaps, the nodal gap structures could be complex and band dependent.

The superconductivity in $CsV_3Sb_5$ is further investigated by imaging magnetic vortices. Figs. 2(a) and 2(b) show the vortex maps at $V_b$ =0 mV and 0.5 mV in a 500 × 500 $nm^2$ Sb surface under $B_\perp$ = 500 Oe. The vortices are clearly visible and form perfect Abrikosov lattice; however, a single vortex is spatially anisotropic and reveals a sixfold star-shaped structure. The Abrikosov lattice is along the crystalline lattice direction, while the star-shaped tails at $V_b$ = 0 mV are rotated by 30° away from the lattice direction. Since the vortex structure is strongly related to the Fermi surface structure and the superconducting gap structure, anisotropic vortex core is generally believed to arise from the anisotropic and/or multiband nature of the Fermi surfaces and/or superconducting pairing, as previously reported in $NbSe_2$, $YNi_2B_2C$, LiFeAs [25-29]. In addition to the multiple sets of coherence peaks in the *dI/dV* spectra, the observed sixfold star-shaped vortices in $CsV_3Sb_5$ further confirm a multiband and/or anisotropic superconducting state.

Figs. 2(c) and 2(d) show the evolution of *dI/dV* spectra across a vortex core, taken along the two line cuts #1 and #2 in Fig. 2(a). Far from the vortex core, the superconducting gap is only slightly affected, and it is gradually suppressed upon approaching the vortex core. At a distance of about ± 40 nm away from the vortex core, a pair of broad peaks inside the superconducting gap appears, and their energy separation decreases gradually upon approaching the vortex core. These two peaks are discernible 10 nm away from the vortex core, but merge into a broad asymmetric peak near the vortex center. Similar results have been observed in the other vortices on the Sb and half-Cs surfaces, as shown in Fig. S5 of SM. The evolution of the vortex core states is consistent with the conventional Caroli–de Gennes–Matricon (CdGM) bound states [30]. $CsV_3Sb_5$ possesses multiple Fermi pockets with the Fermi energy ($E_F$) ranging from 50 meV to 700 meV [10]. By using the observed largest superconducting gap $\Delta_3 \sim 0.57$ meV, the estimated upper limit of the energy separation of the CdGM bound states $\delta E = \Delta^2/E_F$ (Ref. [30]) is about 6.5 μeV, thus the CdGM states of different bands cannot be distinguished in our STM study and appear as a broad peak containing hundreds of CdGM states.

To further explore the superconducting gap structure, impurity-induced effects are studied. In general, the response of superconductivity to local impurities depends on the pairing symmetry and the characteristic of the impurities [31]. It is known that for *s*-wave pairing, only magnetic impurities can break the Cooper pair and induce in-gap bound states [32]. However, for sign-changing pairing functions such as *d*-wave and *s*±-wave, it is predicted that nonmagnetic impurities with proper scattering potentials can also induce in-gap states and suppress superconductivity [33–35], which is supported by STM measurements on cuprates [36], NaFeAs [37], and LiFeAs [38].

We investigate the impurity effects of the superconductivity in $CsV_3Sb_5$ by studying the intrinsic defects, as well as artificially introduced impurities on the cleaved surfaces. The insets of Figs. 3(a) and 3(b) show two kinds of hole-like defects located at the Sb site and the V site underneath, which are assigned as a Sb vacancy and a V defect. The superconducting gap is unaffected at both defects and their vicinity, and no in-gap states are observed, as shown in Figs. 3(a) and 3(b). Since no signature of magnetic order or local magnetic moment was observed in $CsV_3Sb_5$ [6,39], these intrinsic defects are considered to be non-magnetic. Similarly, the non-magnetic Cs adatoms on the Sb surface do not affect the superconducting gap (Fig. 3(c)). Besides, non-magnetic Zn atoms and magnetic Cr atoms were evaporated separately onto the sample while it was kept at about 30 K. In the STM image, these atoms appear as bright protrusions (insets of Figs. 3(d) and 3(e)). Assuming that the interaction between the low-temperature adsorbed atoms and the underlying lattice is weak, the impurity atoms are expected to retain their nonmagnetic/magnetic character after adsorption. For Zn adatoms, the superconducting gap remains unchanged at the Zn site and its vicinity. While for the Cr cluster, the superconducting gap is greatly suppressed, and a pair of asymmetric peaks appear inside the gap. These are hallmarks of impurity-induced in-gap states. Away from the Cr cluster, the impurity states are weakened, and the superconducting gap gradually recovers. Overall, the magnetic Cr cluster strongly suppresses the superconductivity of $CsV_3Sb_5$; while the nonmagnetic impurities, including the intrinsic Sb vacancy, V defect, step edges,

Cs and Zn adatoms, all have no influence on the superconductivity of $CsV_3Sb_5$. These intuitively suggest that there is no sign change in the superconducting pairing function of $CsV_3Sb_5$.

The interplay between the multiband superconductivity and CDWs are manifested in the spatial modulation of the DOS for the two kinds of superconducting gaps. Figs. 4(a)-4(d), together with Figs. S6 and S7 in SM, show the DOS maps and the corresponding FFTs at various energies for the V-shaped superconducting gap, taken in the Sb surface shown in Fig. 1(d). Obvious spatial modulations are observed, as clearly resolved in Fig. 4(e) displaying the spatial dependence of the superconducting DOS at E = 0 meV and -0.36 meV, measured along the two lattice directions (line cuts #3 and #4) in Fig. 4(a). Along line cut #3 without unidirectional $4a_0$ CDW, the DOS oscillations exhibit a period of $a_0$ (labeled $q_a$ in $k$-space); the phases are opposite for DOS oscillations at E = -0.36 meV and 0 mV, which is consistent with the picture that a higher coherence peak is concomitant with a lower residual DOS at zero energy. Such antiphase behavior suggests that the observed DOS modulation is intrinsic to the superconducting state, instead of the influence of atomic Bragg oscillations. Along line cut #4 with unidirectional $4a_0$ CDW, the minimal oscillation period is still $a_0$, but the oscillation intensity is strongly influenced by the unidirectional $4a_0$ CDWs and shows additional $4a_0$ and $(4/3)a_0$ periods. Figs. 4(f)-4(h) show the energy dependencies of these DOS modulations, taken along cuts #5-#7 in $k$-space (Fig. 4(b)). $q_a$, $q_{4a}$ and $q_{(4/3)a}$ modulations are clearly observed, they are non-dispersive and show strong spatial anisotropy. The DOS modulations are very weak for $|E| \sim \Delta_1$, but are much more evident for $|E| \leq \Delta_2$, implying that the unidirectional $4a_0$ CDW coexists with the superconducting gap $\Delta_2$ but not $\Delta_1$. As for the U-shaped superconducting gap, although the $2a_0 \times 2a_0$ and unidirectional $4a_0$ CDWs are also present in the STM image (Fig. 4(i)), the overall superconducting DOS modulation is very weak and the influence of CDWs is not obvious except for those near the coherence peak energies (Figs. 4(j)-4(l) and Fig. S8 in SM). These discrepancies indicate that the interactions between CDWs and different superconducting gaps are different.

$CsV_3Sb_5$ is a multiband system, containing a circular-like electron pocket contributed by Sb-$p$ orbital at $\Gamma$ point, multiple triangle-like electron pockets at $K$ points and multiple saddle points with high DOS at $M$ points originating from V-$d$ orbitals [10,22]. As suggested by theory and recent ARPES studies [40,41], the scattering of electrons among the saddle points at $M$ point drives the $2a_0 \times 2a_0$ CDW order in $CsV_3Sb_5$, and the opening of the CDW gap is strongly momentum dependent. At low temperature,

CDW gap appears around the *M* point and the electronic states remain gapless around the $\Gamma$ point and along the $\Gamma$-*K* direction [41]. The two kinds of superconducting gaps herein are considered to originate from different bands or Fermi pockets, and the different influence of the CDWs on these Fermi pockets will affect their superconducting pairing behavior and result in different superconducting gap functions. The V-shaped superconducting gap, coexisting with CDWs as shown in Fig. 4, probably originates from the anisotropic electron pockets at *K* points; its competition with CDWs is stronger, resulting in a V-shaped gap structure with large residual DOS. The U-shaped gap is more likely derived from the circular electron pocket at $\Gamma$ point, which is less affected by the CDW orders.

In summary, we systematically studied the superconducting properties of CsV$_3$Sb$_5$ by using an ultralow-temperature STM with high energy and spatial resolution. Some direct evidence and critical clues related to the superconducting pairing symmetry are provided, such as the existence of gap nodes, the absence of sign change in the order parameter, multiple superconductivity and its band-dependent interactions with various CDW orders. Our results are thus informative for further revealing the complex entanglements of superconductivity, topology and charge order in kagome compounds.

## ACKNOWLEDGMENTS

We thank Yilin Wang, Juan Jiang and Tong Zhang for helpful discussions. The measurements of magnetic and electrical properties were conducted on the $^3$He insert of MPMS3 and PPMS at Instruments Center for Physical Science of University of Science and Technology of China. This work is supported by the National Natural Science Foundation of China (Grants No. 12074363, No. 11790312, No. 11888101, No.11774060 and No. 92065201), National Key R&D Program of the MOST of China (Grants No. 2017YFA0303004, No. 2016YFA0300200, and No. 2017YFA0303104).

## Author contributions

CsV$_3$Sb$_5$ crystals were grown by Wei Xia and Yanfeng Guo. STM measurements were performed by Han-Shu Xu, Ziyuan Chen, Ruotong Yin, Yuanji Li and Ya-Jun Yan. The data analysis was performed by Han-Shu Xu, Ya-Jun Yan, Shijie Fang, Ruotong Yin and Wenqi Yang. Han-Shu Xu, Ya-Jun Yan and Dong-Lai Feng coordinated the whole work and wrote the manuscript. All authors have discussed the results.

**References**:

[1] L. D. Ye, M. G. Kang, J. W. Liu, F. V. Cube, C. R. Wicker, T. Suzuki, C. Jozwiak, A. Bostwick, E. Rotenberg, D. C. Bell, L. Fu, R. Comin, and J. G. Checkelsky, Massive Dirac fermions in a ferromagnetic kagome metal, *Nature* **555**, 638 (2018).

[2] J.-X. Yin, S. S. Zhang, G. Q. Chang, Q. Wang, S. S. Tsirkin, Z. Guguchia, B. Lian, H. B. Zhou, K. Jiang, I. Belopolski, N. Shumiya, D. Multer, M. Litskevich, T. A. Cochran, H. Lin, Z. Q. Wang, T. Neupert, S. Jia, H. C. Lei, and M. Z. Hasan, Negative flat band magnetism in a spin–orbit-coupled correlated kagome magnet, *Nat. Phys.* **15**, 443 (2019).

[3] L. Balents, Spin liquids in frustrated magnets, *Nature* **464**, 199 (2010).

[4] W.-H. Ko, P. A. Lee, and X.-G. Wen, Doped kagome system as exotic superconductor, *Phys. Rev. B* **79**, 214502 (2009).

[5] W.-S. Wang, Z.-Z. Li, Y.-Y. Xiang, and Q.-H. Wang, Competing electronic orders on kagome lattices at van Hove filling, *Phys. Rev. B* **87**, 115135 (2013).

[6] B. R. Ortiz, L. C. Gomes, J. R. Morey, M. Winiarski, M. Bordelon, J. S. Mangum, I. W. H. Oswald, J. A. Rodriguez-Rivera, J. R. Neilson, S. D. Wilson, E. Ertekin, T. M. McQueen, and E. S. Toberer, New kagome prototype materials: discovery of $KV_3Sb_5$, $RbV_3Sb_5$, and $CsV_3Sb_5$, *Phys. Rev. Materials* **3**, 094407 (2019).

[7] B. R. Ortiz, S. M. L. Teicher, Y. Hu, J. L. Zuo, P. M. Sarte, E. C. Schueller, A. M. M. Abeykoon, M. J. Krogstad, S. Rosenkranz, R. Osborn, R. Seshadri, L. Balents, J. F. He, and S. D. Wilson, $CsV_3Sb_5$: A $\mathbb{Z}_2$ Topological Kagome Metal with a Superconducting Ground State, *Phys. Rev. Lett.* **125**, 247002 (2020).

[8] B. R. Ortiz, P. M. Sarte, E. M. Kenney, M. J. Graf, S. M. L. Teicher, R. Seshadri, and S. D. Wilson, Superconductivity in the $\mathbb{Z}_2$ kagome metal $KV_3Sb_5$, *Phys. Rev. Materials* **5**, 034801 (2021).

[9] Q. W. Yin, Z. J. Tu, C. S. Gong, Y. Fu, S. H. Yan and H. C. Lei, Superconductivity and Normal-State Properties of Kagome Metal $RbV_3Sb_5$ Single Crystals, *Chin. Phys. Lett.* **38**, 037403 (2021).

[10] Z. H. Liu, N. N. Zhao, Q. W. Yin, C. S. Gong, Z. J. Tu, M. Li, W. H. Song, Z. T. Liu, D. W. Shen, Y. B. Huang, K. Liu, H. C. Lei, and S. C. Wang, Temperature-induced band renormalization and Lifshitz transition in a kagome superconductor $RbV_3Sb_5$, *arXiv*:2104.01125 (2021).

[11] Y.-X. Jiang, J.-X. Yin, M. M. Denner, N. Shumiya, B. R. Ortiz, J. Y. He, X. X. Liu, S. S. Zhang, G. Q. Chang, I. Belopolski, Q. Zhang, M. S. Hossain, T. A. Cochran, D. Multer, M. Litskevich, Z.-J. Cheng, X. P. Yang, Z. Guguchia, G. Xu, Z. Q. Wang, T. Neupert, S. D. Wilson, M. Z. Hasan, Discovery of topological charge order in kagome superconductor $KV_3Sb_5$, *arXiv*:2012.15709 (2020).

[12] S.-Y. Yang, Y. Wang, B. R. Ortiz, D. Liu, J. Gayles, E. Derunova, R. Gonzalez-Hernandez, L. Šmejkal, Y. L. Chen, S. S. P. Parkin, S. D. Wilson, E. S. Toberer, T. McQueen and M. N. Ali, Giant, unconventional anomalous Hall effect in the metallic frustrated magnet candidate $KV_3Sb_5$, *Sci. Adv.* **6**,

eabb6003 (2020).

[13] F. H. Yu, T. Wu, Z. Y. Wang, B. Lei, W. Z. Zhuo, J. J. Ying, and X. H. Chen, Concurrence of anomalous Hall effect and charge density wave in a superconducting topological kagome metal, *arXiv*:2102.10987 (2021).

[14] Y. J. Wang, S. Y. Yang, P. K. Sivakumar, B. R. Ortiz, S. M. L. Teicher, H. Wu, A. K. Srivastava, C. Garg, D. F. Liu, S. S. P. Parkin, E. S. Toberer, T. McQueen, S. D. Wilson, and M. N. Ali, Proximity-induced spin-triplet superconductivity and edge supercurrent in the topological Kagome metal, $K_{1-x}V_3Sb_5$, *arXiv*:2012.05898 (2020).

[15] C. C. Zhao, L. S. Wang, W. Xia, Q. W. Yin, J. M. Ni, Y. Y. Huang, C. P. Tu, Z. C. Tao, Z. J. Tu, C. S. Gong, H. C. Lei, Y. F. Guo, X. F. Yang, and S. Y. Li, Nodal superconductivity and superconducting domes in the topological Kagome metal $CsV_3Sb_5$, *arXiv*:2102.08356 (2021).

[16] K. Y. Chen, N. N. Wang, Q. W. Yin, Z. J. Tu, C. S. Gong, J. P. Sun, H. C. Lei, Y. Uwatoko, and J.-G. Cheng, Double superconducting dome and triple enhancement of Tc in the kagome superconductor $CsV_3Sb_5$ under high pressure, *arXiv*:2102.09328 (2021).

[17] F. Du, S. S. Luo, B. R. Ortiz, Y. Chen, W. Y. Duan, D. T. Zhang, X. Lu, S. D. Wilson, Y. Song, and H. Q. Yuan, Pressure-tuned interplay between charge order and superconductivity in the kagome metal $KV_3Sb_5$, *arXiv*:2102.10959 (2021).

[18] Z. Y. Zhang, Z. Chen, Y. Zhou, Y. F. Yuan, S. Y. Wang, L. L. Zhang, X. D. Zhu, Y. H. Zhou, X. L. Chen, J. H. Zhou, and Z. R. Yang, Pressure-induced Reemergence of Superconductivity in Topological Kagome Metal $CsV_3Sb_5$, *arXiv*:2103.12507 (2021).

[19] W. Y. Duan, Z. Y. Nie, S. S. Luo, F. H. Yu, B. R. Ortiz, L. C. Yin, H. Su, F. Du, A. Wang, Y. Chen, X. Lu, J. J. Ying, S. D. Wilson, X. H. Chen, Y. Song, and H. Q. Yuan, Nodeless superconductivity in the kagome metal $CsV_3Sb_5$, *arXiv*:2103.11796 (2021).

[20] Z. W. Liang, X. Y. Hou, W. R. Ma, F. Zhang, P. Wu, Z. Y. Zhang, F. H. Yu, J.-J. Ying, K. Jiang, L. Shan, Z. Y. Wang, and X.-H. Chen, Three-dimensional charge density wave and robust zero-bias conductance peak inside the superconducting vortex core of a kagome superconductor $CsV_3Sb_5$, *arXiv*:2103.04760 (2021).

[21] Supplemental Materials for Figs. S1–S8.

[22] H. Zhao, H. Li, B. R. Ortiz, S. M. L. Teicher, T. Park, M. X. Ye, Z. Q. Wang, L. Balents, S. D. Wilson, and I. Zeljkovic, Cascade of correlated electron states in a kagome superconductor $CsV_3Sb_5$, *arXiv*:2103.03118 (2021).

[23] H. Chen, H. T. Yang, B. Hu, Z. Zhao, J. Yuan, Y. Q. Xing, G. J. Qian, Z. H. Huang, G. Li, Y. H. Ye, Q. W. Yin, C. S. Gong, Z. J. Tu, H. C. Lei, S. Ma, H. Zhang, S. L. Ni, H. X. Tan, C. M. Shen, X. L. Dong, B. H. Yan, Z. Q. Wang, and H.-J. Gao, Roton pair density wave and unconventional strong-coupling superconductivity in a topological kagome metal, *arXiv*:2103.09188 (2021).

[24] H. X. Tan, Y. Z. Liu, Z. Q. Wang, and B. H. Yan, Charge density waves and electronic properties

of superconducting kagome metals, *arXiv*:2103.06325 (2021).

[25] H. F. Hess, R. B. Robinson, and J. V. Waszczak, Vortex-Core Structure Observed with a Scanning Tunneling Microscope, *Phys. Rev. Lett.* **64**, 2711 (1990).

[26] H. Nishimori, K. Uchiyama, S.-i. Kaneko, A. Tokura, H. Takeya, K. Hirata, and N. Nishida, First Observation of the Fourfold-symmetric and Quantum Regime Vortex Core in $YNi_2B_2C$ by Scanning Tunneling Microscopy and Spectroscopy, *J. Phys. Soc. Jpn.* **73**, 12 (2004).

[27] T. Hanaguri, K. Kitagawa, K. Matsubayashi, Y. Mazaki, Y. Uwatoko, and H. Takagi, Scanning tunneling microscopy/spectroscopy of vortices in LiFeAs, *Phys. Rev. B* **85**, 214505 (2012).

[28] N. Hayashi, M. Ichioka, and K. Machida, Star-Shaped Local Density of States around Vortices in a Type-II Superconductor, *Phys. Rev. Lett.* **77**, 4074 (1996).

[29] N. Hayashi, M. Ichioka, and K. Machida, Effects of gap anisotropy upon the electronic structure around a superconducting vortex, *Phys. Rev. B* **56**, 9052 (1997).

[30] C. Caroli, P. G. De Gennes, and J. Matricon, Bound Fermion States on a Vortex Line in a Type II Superconductor, *Phys. Lett.* **9**, 307 (1964).

[31] A. V. Balatsky, I. Vekhter, and J.-X. Zhu, Impurity-induced states in conventional and unconventional superconductors, *Rev. Mod. Phys.* **78**, 373 (2006).

[32] P. W. Anderson, Theory of dirty superconductors, *J. Phys. Chem. Solids* **11**, 26 (1959).

[33] S. Onari and H. Kontani, Violation of Anderson's Theorem for the Sign-Reversing *s*-Wave State of Iron-Pnictide Superconductors, *Phys. Rev. Lett.* **103**, 177001 (2009).

[34] D. Zhang, Nonmagnetic Impurity Resonances as a Signature of Sign-Reversal Pairing in FeAs-Based Superconductors, *Phys. Rev. Lett.* **103**, 186402 (2009).

[35] T. Kariyado and M. Ogata, Single-Impurity Problem in Iron-Pnictide Superconductors, *J. Phys. Soc. Jpn.* **79**, 083704 (2010).

[36] S. H. Pan, E. W. Hudson, K. M. Lang, H. Eisaki, S. Uchida, and J. C. Davis, Imaging the effects of individual zinc impurity atoms on superconductivity in $Bi_2Sr_2CaCu_2O_{8+\delta}$, *Nature* **403**, 746 (2000).

[37] H. Yang, Z. Y Wang, D. L. Fang, Q. Deng, Q. H. Wang, Y. Y. Xiang, Y. Yang, and H. H. Wen, In-gap quasiparticle excitations induced by non-magnetic Cu impurities in $Na(Fe_{0.96}Co_{0.03}Cu_{0.01})As$ revealed by scanning tunnelling spectroscopy, *Nat. Commun.* **4**, 2749 (2013).

[38] S. Grothe, S. Chi, P. Dosanjh, R. X. Liang, W. N. Hardy, S. A. Burke, D. A. Bonn, and Y. Pennec, Bound states of defects in superconducting LiFeAs studied by scanning tunneling spectroscopy, *Phys. Rev. B* **86**, 174503 (2012).

[39] E. M. Kenney, B. R. Ortiz, C. N. Wang, S. D. Wilson, and M. J. Graf, Absence of local moments in the kagome metal $KV_3Sb_5$ as determined by muon spin spectroscopy, *J. Phys.: Condens. Matter* (in press) https://doi.org/10.1088/1361-648X/abe8f9 (2021).

[40] M. M. Denner, R. Thomale, and T. Neupert, Analysis of charge order in the kagome metal $AV_3Sb_5$ (A=K, Rb, Cs), *arXiv*:2103.14045 (2021).

[41] Z. G. Wang, S. Ma, Y. H. Zhang, H. T. Yang, Z. Zhao, Y. Ou, Y. Zhu, S. L. Ni, Z. Y. W. Lu, H. Chen, K. Jiang, L. Yu, Y. Zhang, X. L. Dong, J. P. Hu, H.-J. Gao, and Z. X. Zhao, Distinctive momentum dependent charge-density-wave gap observed in $CsV_3Sb_5$ superconductor with topological Kagome lattice, *arXiv*: 2104.05556 (2021).

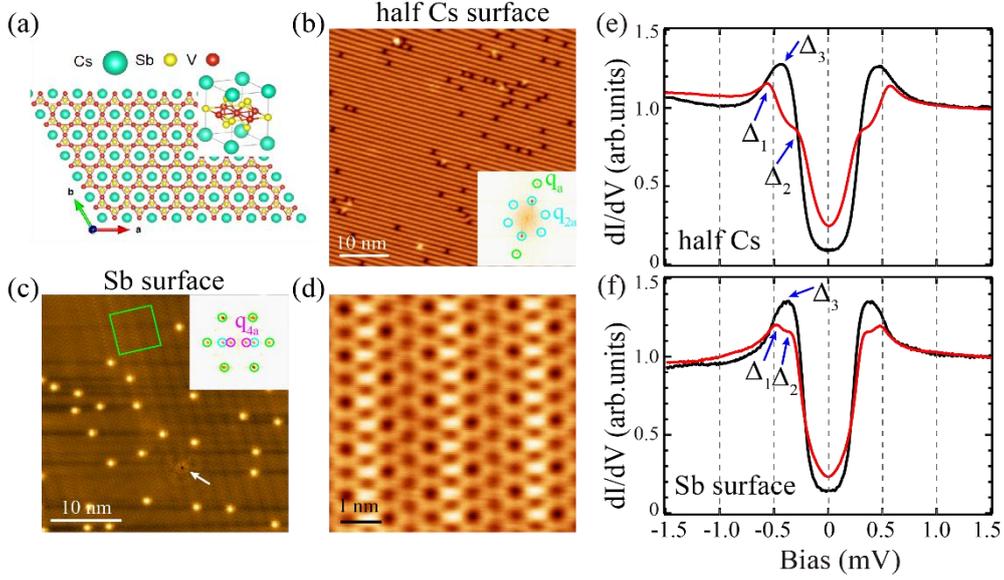

FIG. 1. STM images and superconducting spectra of $CsV_3Sb_5$. (a) The prototype structure of $CsV_3Sb_5$, exhibiting a layered structure of V-Sb sheets intercalated by Cs layers. (b)(c) Typical STM images of the half Cs surface (50×50 nm$^2$, $V_b$ = 200 mV, $I_t$ = 60 pA) and the Sb surface (35×35 nm$^2$, $V_b$ = 20 mV, $I_t$ = 60 pA), with corresponding FFTs shown in the insets. The Bragg peak $q_a$, $2a_0×2a_0$ CDW $q_{2a}$ and unidirectional modulation $q_{4a}$, are marked by green, cyan and magenta circles, respectively. (d) Atomically resolved STM image of the Sb surface taken in the green box in (c) (6×6 nm$^2$, $V_b$ = 70 mV, $I_t$ = 209 pA). (e)(f) Two kinds of superconducting gap spectra observed on these two surfaces ($V_b$ = 1.2 mV, $I_t$ = 150 pA, $\Delta V$ = 30 μV).

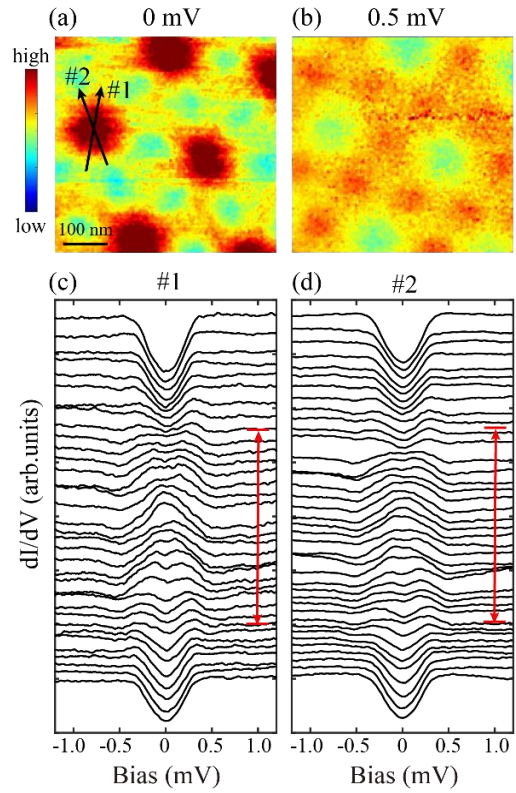

FIG. 2. Vortex states on the Sb surface. (a)(b) Vortex maps at $V_b = 0$ mV and 0.5 mV under $B_\perp = 500$ Oe (500 × 500 nm$^2$, $V_b = 1.20$ mV, $I_t = 100$ pA, $\Delta V = 30$ μV). (c)(d) Evolution of the $dI/dV$ spectra taken along the two line cuts #1 and #2 indicated in (a) ($V_b = 1.20$ mV, $I_t = 150$ pA, $\Delta V = 30$ μV). The red bidirectional arrows indicate the ± 40 nm area away from the vortex core.

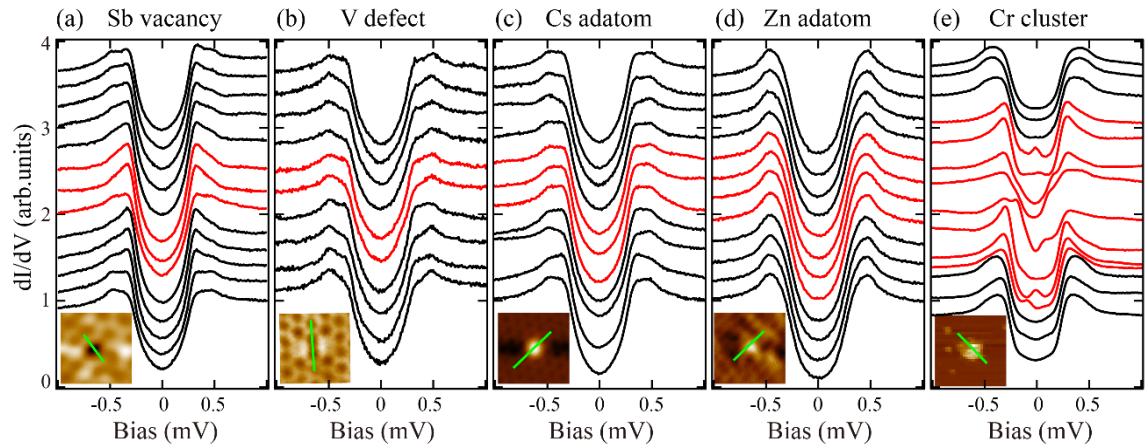

FIG.3. Impurity-induced effects on the superconductivity of $CsV_3Sb_5$. Series of *dI/dV* spectra taken along the line cuts (green lines in the insets) across the intrinsic defects ((a) a Sb-site vacancy and (b) a V-site defect) as well as surface adsorbed atoms/clusters ((c) a Cs adatom, (d) a Zn adatom, (e) a Cr cluster). The corresponding topographic images of these defects/adatoms are displayed in the insets. The red curves are collected near the center of the defects/adatoms.

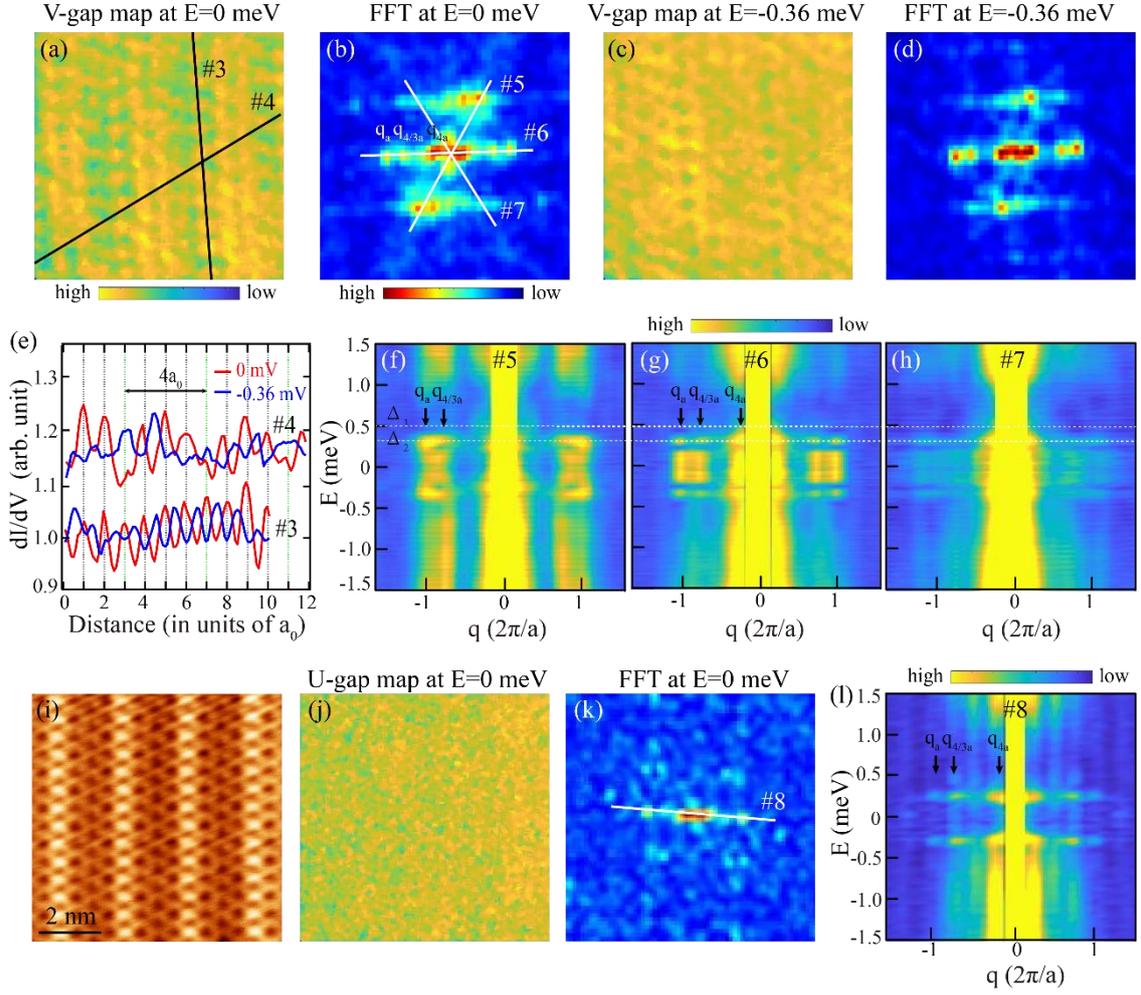

FIG. 4. Spatial modulation of the superconducting DOS for the V-shaped (a-h) and U-shaped (i-l) superconducting gaps on the Sb surface. (a-d) DOS maps and the corresponding FFTs under $E = 0$ meV and -0.36 meV, taken in the Sb surface shown in Fig. 1(b) that exhibiting a V-shaped superconducting gap. Measurement conditions: 6×6 nm$^2$, $V_b = 1.20$ mV, $I_t = 300$ pA, $\Delta V = 30$ μV. (e) Comparison of spatial modulations of the DOS under $E = 0$ meV and -0.36 meV, taken along the line cuts #3 and #4 in (a), respectively. (f-h) Color plots of FFT line cuts along #5, #6 and #7 directions in (b). The periods of $q_a$, $q_{4a/3}$ and $q_{4a}$ are marked. Energy positions of $\Delta_1$ and $\Delta_2$ are indicated by the white dashed lines. (i) Atomically resolved STM image of the Sb surface with a U-shaped superconducting gap (8×8 nm$^2$, $V_b = 50$ mV, $I_t = 20$ pA). (j)(k) DOS map and the corresponding FFT under $E = 0$ meV taken in the area shown in (i) ($V_b = 1.20$ mV, $I_t = 200$ pA, $\Delta V = 30$ μV.). (l) Color plot of FFT line cut along #8 direction in (k).